\documentclass[preprint,showpacs,preprintnumbers,amsmath,amssymb]{revtex4}

\usepackage{graphicx}% Include figure files
\usepackage{dcolumn}% Align table columns on decimal point
\usepackage{bm}% bold math

\begin{document}

\title{Low-energy photodisintegration of $^9$Be \\ 
with the molecular orbit model}

\author{N. Itagaki$^1$}
 \email{itagaki@phys.s.u-tokyo.ac.jp}
\author{K. Hagino$^2$}
 \email{hagino@yukawa.kyoto-u.ac.jp}
 \affiliation{$^1$Department of physics, University of Tokyo, Hongo, 
Tokyo 113-0033, Japan \\
$^2$Yukawa Institute for Theoretical Physics, Kyoto
University, Kyoto 606-8502, Japan }

\date{\today}% It is always \today, today,

\begin{abstract}

We use a microscopic three-cluster model with the $\alpha+\alpha+n$
configuration to 
analyze with astrophysical interests 
the photodisintegration cross section of $^9$Be.  
The valence neutron
in $^9$Be is treated in the molecular orbit model (MO), including 
continuum states by the box
discretization method. 
It is shown that good agreement
is achieved with the recently measured B(E1) transition probability
from the ground state to the first 1/2$^+$ state when
the box size is enlarged, 
resulting in reasonable reproduction of the
experimental data for the photodisintegration cross section. 
The Coulomb dissociation of $^9$Be is also discussed. 

\end{abstract}

\maketitle

The $^9$Be nucleus has been popular especially 
for the past few years.
Firstly, it has been often used as a target for projectile fragmentation
reactions which involve the radioactive beams \cite{Navin00}.
In order to analyze the experimental data, a new neutron-$^9$Be
optical potential was also constructed \cite{BB01}.
Secondly, the nucleus
has been utilized to investigate the role of break-up process
in heavy-ion subbarrier fusion reactions \cite{DHB99,HDF02}.
It was reported in Ref. \cite{DHB99} that the complete fusion cross
sections for the $^9$Be + $^{208}$Pb reaction
were hindered by 68\% at energies above the Coulomb barrier compared with
those expected from a simple one dimensional potential model.
This large suppression of complete fusion cross sections was
attributed to the low energy thresholds for breakup of $^9$Be into
charged fragments, i.e.,
$Q=-1.57$ MeV for $^9$Be $\to \alpha + \alpha + n$ and
$Q=-2.47$ MeV for $^9$Be $\to \alpha + ^5$He.
The subbarrier breakup reaction of $^9$Be
was also measured subsequently, observing large $\alpha$ particle
cross sections even at energies well below the Coulomb barrier \cite{HDF02}.

$^9$Be is interesting also from the nuclear structure point of
view \cite{OA79,D89,AOSV96,EH01,IO00}.
The prominent $\alpha$ cluster structure is well established for this
nucleus, and furthermore, it is a loosely bound nucleus in a sense that 
none of the two-body subsystems 
($\alpha$-$\alpha$ or $\alpha$-$n$) is particle stable.
These properties make $^9$Be a good testing ground for any nuclear structure
model applying to heavier Be isotopes.
Based on a similar idea as the linear combination of atomic orbitals
(LCAO) for hydrogen molecules in quantum chemistry, the molecular orbit
model (MO) was developed for the $^9$Be nucleus\cite{OA79}. This model
has recently 
been applied with success to other heavier isotopes as well
\cite{IO00,IOI00,IOI01,IHOOI02}.
For instance, it was pointed out that the presence of valence neutrons
enhances the $\alpha$ cluster structure of $^{10}$Be 
by occupying the $1/2^+$ ($\sigma$-) orbit in the $sd$-shell \cite{IO00}. 

Another important aspect of the $^9$Be nucleus is its role in
the $r$-process
nucleosynthesis for heavier elements in the neutrino-driven wind formed
in core-collapse supernovae. In such an exploding environment with
abundant neutrons,
the three-body reaction $\alpha(\alpha n,\gamma)^9$Be, followed by
$^9$Be($\alpha,n)^{12}$C, plays a key role to bridge the unstable mass
gaps at $A$=5 and 8 compared with the triple-$\alpha$
reaction \cite{WH92,SUGK02}.  
The reaction rate for the former
process is governed by the low energy resonances of $^9$Be, which can
be determined experimentally by measuring the cross section for the
inverse reaction, i.e., the photodisintegration
process $^9$Be($\gamma,\alpha n)\alpha$ \cite{SUGK02}.
Utsunomiya {\it et al.} recently performed such measurement using
laser-induced Compton backscattered $\gamma$ rays \cite{UYA00}.
They obtained the B(E1) transition probability 
from the ground $3/2^-$ state
to the first 1/2$^+$ state (the first excited state) 
almost twice as large as the previous measurement 
which used inelastic electron scattering, and
the resonance energy of the first 1/2$^+$ state which
is slightly shifted upperward by 0.064 MeV \cite{UYA00}.

From the theoretical point of view, 
the treatment of the $1/2^+$ state in $^9$Be  
has been a challenge, since it 
is not bound but lies above the
$^8$Be+$n$ threshold only by 19 keV ($E_x = 1.68$ MeV). 
The aim of this paper is to 
present a detailed analysis of this state based on the 
molecular orbit method, and apply it 
to the photodisintegration of $^9$Be in order to analyze the new experimental
data of Utsunomiya {\it et al.}.
Although Descouvemont has done similar studies based on a microscopic
three-cluster model and obtained nice agreement with the
experimental data for the B(E1) value as well as the
photodisintegration cross sections \cite{D01},
we particularly focus on the dependence of the B(E1)
value on the properties of the wave function for the valence neutron.
Therefore, to some extent,
our study is complementary to Ref. \cite{D01}.
In addition, we also give a brief discussion on the subbarrier Coulomb
dissociation of $^9$Be.

We introduce a microscopic $\alpha$+$\alpha$+$n$ model for $^9$Be. 
We first consider 
an intrinsic wave function as a basis where 
two alpha particles and the valence 
neutron are located at $\pm \vec{d}/2$ and $\vec{R}_n$ 
against the origin, respectively. 
The total intrinsic wave function is fully
antisymmetrized, and 
the single particle wave function for all the nucleons is expressed by a 
Gaussian function. 
Each $\alpha$ cluster contains four nucleons, and its wave function 
reads
\begin{eqnarray}
&&\phi^{(\alpha)}(\vec{r}_{p\uparrow},\vec{r}_{p\downarrow},
\vec{r}_{n\uparrow},\vec{r}_{n\downarrow}) \nonumber \\
&&=G_{\vec R_\alpha}(\vec{r}_{p\uparrow})
G_{\vec R_\alpha}(\vec{r}_{p\downarrow})
G_{\vec R_\alpha}(\vec{r}_{n\uparrow})
G_{\vec R_\alpha}(\vec{r}_{n\downarrow})
\chi_{p\uparrow}
\chi_{p\downarrow}
\chi_{n\uparrow}
\chi_{n\downarrow}, 
\end{eqnarray}
where $\chi$ denotes the spin-isospin wave function.  
Here, $G$ represents a Gaussian function, 
\begin{equation}
G_{\vec R}(\vec{r}) = \frac{1}{( \pi s^2)^{3/4}}
\exp[-(\vec r-\vec R)^2/2s^2],
\label{gauss}
\end{equation}
and $\vec R_\alpha = \pm \vec{d}/2$. 
The wave function for the 
valence neutron 
is also expressed by a Gaussian (\ref{gauss}) with 
$\vec{R}=\vec{R}_n$. 
The total intrinsic wave function is then projected onto 
the eigen-states of angular momentum $J$ and parity, 
which we perform numerically. 
Note that for $^9$Be 
the MO picture can be introduced 
by merely performing the parity projection, 
since the $\alpha$-$\alpha$ core is always an even-parity system.

The spectra of $^9$Be are obtained by 
superposing the basis states
with different $\alpha$-$\alpha$ distances $d$ as well as 
neutron distances $\vec{R}_n$. The coefficients for the superposition 
are determined by the variational principle. 
To this end, we descretize $\vec R_n$, and include from zero to
$|\vec R_n^{\rm max}| =15$fm with a step size of typically 2 fm. 
The $\alpha$-$\alpha$ distances $d$ and the relative angle between 
$\vec{d}$ and $\vec{R}_n$ are also discretized similarly. 
For the oscillator parameter in the Gaussian function (\ref{gauss}), 
we use $s$ = 1.46 fm in the following calculations.
The adopted effective nucleon-nucleon interaction 
is the Volkov No.2 
with the exchange parameters $M = 0.6$ ($W = 0.4$)
for the central part,
and the G3RS spin-orbit term 
for the spin-orbit part as in Ref. \cite{IO00}.
All the parameters of this interaction were determined
from the $\alpha+n$ and $\alpha+\alpha$
scattering phase shifts.

Our interest in this paper 
is to compute the photo reaction cross sections of 
$^9$Be. For this purpose, 
we assume that the $^9$Be eventually breaks up to
$\alpha + \alpha + n$ with
100\% of probability once it is excited above the threshold for
neutron emission.
With this assumption, the photodisintegration cross sections are
equivalent to the photo absorption cross sections.
Since the calculations are based on the bound state approximation,
the continuum states appear as discrete states.
For unpolarized photons, using the Fermi's Golden rule, the E1
photoabsorption
probability per unit time from a bound state $i$ to $f$
reads \cite{JJS}
\begin{equation}
w^{(\rm E1)}_{i\to f} (E_\gamma)=
\frac{e^2}{2\pi}\,\frac{1}{3}\,\frac{E_\gamma}{\hbar}\,
|\langle f|\vec{r}|i\rangle|^2 \,
\delta(E_f-E_i-E_\gamma),
\end{equation}
where $E_\gamma$ is the photon energy, and $E_i$ and $E_f$ is the
energy of the state $i$ and $f$, respectively. The photo absorption
cross sections are obtained by dividing the probability
$w^{(\rm E1)}_{i\to f}$
by the photon flux, $c/(2\pi)^3$,
\begin{equation}
\sigma^{(\rm E1)}_{i\to f} (E_\gamma)=
\frac{16\pi^3}{9}\,\frac{E_{\gamma}}{\hbar c}\,
B({\rm E1};i\to f)\,\delta(E_f-E_i-E_{\gamma}).
\label{photo}
\end{equation}
Here we have carried out the summation for different $m$ substates for
the states $i$ and $f$.
In order to mimic the neutron emission from the excited states of
$^9$Be, we smear the delta function in Eq. (\ref{photo}) by
the Breit-Wigner function as
\begin{equation}
\sigma^{(\rm E1)}_{i\to f} (E_\gamma)=
\frac{16\pi^3}{9}\,\frac{E_{\gamma}}{\hbar c}\,
B({\rm E1};i\to f)\,
\frac{\Gamma/2\pi}{(E_\gamma-E_f+E_i)^2+\Gamma^2/4}.
\label{bw}
\end{equation}
The total cross sections are then obtained by summing up all the final
states $f$.

In the present model, the energy of the ground $3/2^-$ state of $^9$Be
and the ground $0^+$ state of $^8$Be is calculated as 
$-56.3$ MeV and $-54.7$ MeV, respectively. Since 
it is experimentally known that the $^9$Be
nucleus is bound from the $^8$Be+$n$ threshold by 1.7 MeV,
the present model provides the correct binding energy for the
ground state of $^9$Be within 0.1 MeV accuracy, which 
is remarkably good.  
The convergence feature of the energy of the ground state 
as well as the unbound first $1/2^+$ state 
(just above the $^8$Be+$n$ threshold) with respect to the 
maximum spatial separation $|\vec R_n^{\rm max}|$ for the valence neutron 
is listed in Table I, together with the corresponding B(E1) values.
Here, the Gaussian center of 
the valence neutron is incorporated up to
(I): 2.5 fm, (II): 6.0 fm, (III): 10.0 fm, and (IV): 15.0 fm,
measured from the center of mass of the $\alpha$-$\alpha$ core,
respectively.
Since the ground $3/2^-$ state is bound from the threshold by only 1.7
MeV and it thus has a rather long tail, 
we see that at least $|\vec R_n^{\rm max}|$ 
= 10.0 fm is needed for the quantitative
description of the ground state of $^9$Be. 
Note that $|\vec R_n^{\rm max}|$ = 4 fm was employed 
in the previous MO calculation for $^9$Be \cite{IO00}. 
For the unbound $1/2^+$ state, 
the main component is the $s$-wave, which is free from the centrifugal 
barrier, and an extended model space is required.
We find that the model space (IV), where the wave function of the
valence neutron
is solved up to 15 fm from the core, is needed to reach the convergence.

\begin{table}
\caption{The energy convergence of the ground $3/2^-$ state
and the first $1/2^+$ state of $^9$Be. The Gaussian center of 
the valence neutron is incorporated up to $| \vec R_n^{\rm max}| $=
(I): 2.5 fm, (II): 6.0 fm, (III): 10.0 fm, and (IV): 15.0 fm,
measured from the center of mass of the $\alpha$-$\alpha$ core,
respectively.
The energy of the $0^+$ ground state of the 
$\alpha$-$\alpha$ core is calculated to be $-54.7$ MeV.
The B(E1) value from the ground $3/2^-$ state to the first $1/2^+$ state 
is also listed for each model space. 
}

\begin{center}
\begin{tabular}{|c|c|c|c|c|}
\hline
\hline
      &  $| \vec R_n^{\rm max}|$ (fm) &
$3/2^-$ (MeV) & $1/2^+$ (MeV) & B(E1) ($e^2$ fm$^2$)$\uparrow$ \\
\hline
(I)   & 2.5 & $-54.5$ & $-50.3$ & 0.022 \\
(II)  & 6.0 & $-55.8$ & $-53.1$ & 0.032 \\
(III) & 10.0& $-56.3$ & $-54.0$ & 0.042 \\
(IV)  & 15.0& $-56.3$ & $-54.3$ & 0.048 \\
\hline
\hline
\end{tabular}
\end{center}
\end{table}

For the $\alpha+\alpha+n$ system, 
the E1 transition is allowed only as a consequence of the recoil 
effect of the $\alpha$-$\alpha$ core
caused by the valence neutron around it, since we do not
introduce the effective charge.
Notice that 
the B(E1) value is exactly zero for the $\alpha$-$\alpha$ core
without the valence neutron. 
Therefore, 
it is crucial to take a large enough model space for 
the valence neutron in order to obtain a 
quantitative prediction for the B(E1) value. 
The calculated 
B(E1) value increases by more than a factor of two when the model space 
is enlarged from (I) to (IV), 
leading to 
a value which is consistent with the recent experiment by Utsunomiya $et\ al.$
(0.054 $\pm$ 0.004 $e^2$ fm$^2$) \cite{UYA00}. 

\begin{table}
\caption{The excitation energy and the E1 transition probability from 
the ground state to the excited states used to obtain the
photodisintegration cross sections. The $(*)$ denotes the resonance
states. Only those states whose excitation energy is below 5 MeV are
shown. }

\begin{center}
\begin{tabular}{|l|c|c|}
\hline
$J^{\pi}$ & $E_x$ (MeV) & B(E1)$\uparrow$ (e$^2$ fm$^2$)\\
\hline
1/2$^+$ (*)   &  2.05 &     0.048 \\
1/2$^+$   &  3.55  & 0.0062\\
3/2$^+$ (*)  &  4.42   & 0.0088\\
3/2$^+$   &  4.91  & 0.018\\
5/2$^+$ (*)  &  3.13  & 0.028\\
5/2$^+$   &  4.93  & 0.035\\
\hline
\hline
\end{tabular}
\end{center}
\end{table}

\begin{figure}
%  \begin{center}
%    \leavevmode
%    \parbox{0.9\textwidth}
%           {\psfig{file=photo.eps,width=0.8\textwidth}}
%  \end{center}

\includegraphics[width=8cm]{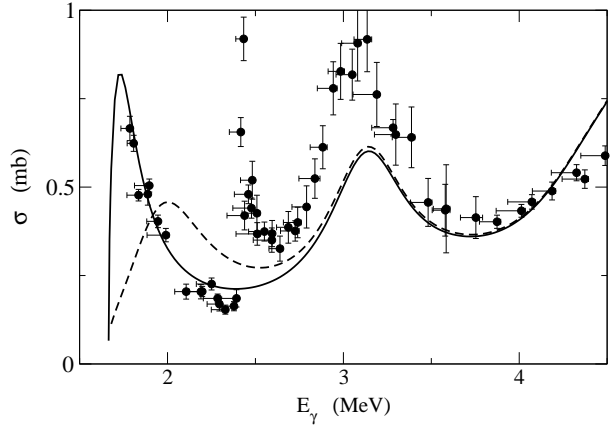}% 
\protect\caption{
The E1 photodisintegration cross sections of $^9$Be as a function of photon
energy obtained in the molecular orbital model (the dashed line).
The experimental data are taken from Ref. {\protect \cite{UYA00}}.
The solid line is the result of the molecular orbital model but
with use of the experimental energy for the lowest 1/2$^+$ state. }
\label{fig:photo}
\end{figure}

Adopting the B(E1) value for the model space (IV), we show in 
Fig. \ref{fig:photo}
the photodisintegration cross sections of $^9$Be as a function of the 
photon energy $E_\gamma$. 
The experimental data are taken from Ref. \cite{UYA00}.
Table II lists the states included in the calculation, 
which are obtained by the present MO method. 
For the width $\Gamma$ in eq.(\ref{bw}), 
we used the experimental values for the
first 3/2$^+$ and 5/2$^+$ resonance states,
while we arbitrarily chose $\Gamma$ = 1
MeV for non-resonant continuum states.
For the width for the first 1/2$^+$ state, we use the energy dependent
width given in Ref. \cite{UYA00}.
The dashed line shows the results of the present calculation thus obtained.
It appears that the low energy cross sections are sensitive to the
energy of the lowest 1/2$^+$ state as well as the B(E1) value to this state.
We found that the energy of the 1/2$^+$ state was rather
sensitive to the box size, even after the B(E1) value was
converged.
Since we could not obtain the converged energy for the 1/2$^+$
state with controlled numerical accuracy,
we instead used the experimental value for it ($E_x=$1.78 MeV), 
although we used the calculated value for the B(E1) (see the solid line).
Besides this point, we see that
the agreement of the theoretical results with the
experimental cross sections is satisfactory.

We have also applied the present calculation to the Coulomb excitations of
$^9$Be. In Ref. \cite{HDF02}, it was shown that, although the energy
dependence was well reproduced,
the measured probability for the Coulomb excitation for the
$^9$Be + $^{208}$Pb reaction were underestimated by a factor of two by
a calculation which included the first $1/2^+$ and $5/2^-$ states
together with
the experimental values for the excitation energies and the
electromagnetic transition probabilities.
The underestimation of the Coulomb excitation probability is likely
attributed to the lack of non-resonant continuum states in the
calculation. We therefore repeated the calculation by including 
the states shown in Table II. 
It was found, however, that the number of continuum states which we
obtained by the present MO calculation 
was too small to resolve the discrepancy.
This is certainly a limitation of the box discretization, and
one would need the proper scattering boundary condition
for the continuum states in order to reconcile the discrepancy.

In summary,
the photodisintegration cross section of $^9$Be 
newly measured by Utsunomiya $et\ al.$ was successfully reproduced
by a microscopic $\alpha$-cluster model. 
It was found that the B(E1) transition
from the ground state to the first $1/2^+$ state,
which plays an important role in explosive nucleosynthesis,  
has a strong dependence on the model space for the valence neutron. 
Incorporating the neutron orbit up to around 15 fm from the $^8$Be
core, we obtained the B(E1) value 
consistent with the new experimental value, which is almost twice the
previous measurement. 

\bigskip

We thank H. Utsunomiya for useful discussions and providing us with
the experimental data in a numerical form.
K.H. thanks D.J. Hinde and M. Dasgupta for useful discussions.
This work is supported in part by Grant-in-Aid 
for Scientific Research (13740145) from the Ministry of Education,
Science and Culture.

\end{document}